\documentclass[
 reprint,
 aps,
 superscriptaddress,
 preprintnumbers,
 amsmath
 amssymb,
 prx
 floatfix,
]{revtex4-2}

\usepackage{graphicx}
\usepackage{dcolumn}
\usepackage{bm}
\usepackage{makecell}
\usepackage{braket} 
\usepackage{siunitx}
\usepackage{color}
\usepackage{array}
\usepackage{xr}
\usepackage{upgreek}
\usepackage[colorlinks]{hyperref}
\usepackage{amsfonts}
\hypersetup{
    citecolor=blue, 
    linkcolor=blue,
    urlcolor=magenta,
}
\usepackage{multirow} 
\usepackage{amsmath}
\usepackage[caption=false]{subfig}

\begin{document}
\title{Emergent Harmonics in Josephson Tunnel Junctions Due to Series Inductance}
\def\RLEaffil{Research Laboratory of Electronics, Massachusetts Institute of Technology, Cambridge, MA 02139, USA}
\def\LLaffil{Lincoln Laboratory, Massachusetts Institute of Technology, Lexington, MA 02421, USA}
\def\Physaffil{Department of Physics, Massachusetts Institute of Technology, Cambridge, MA 02139, USA}
\def\EECSaffil{Department of Electrical Engineering and Computer Science, Massachusetts Institute of Technology, Cambridge, MA 02139, USA}

\author{Junghyun~Kim}
\email[These authors contributed equally;~]{kimjung@mit.edu}
\affiliation{\RLEaffil}
\affiliation{\EECSaffil}

\author{Max~Hays}
\email[These authors contributed equally;~]{maxhays@mit.edu}
\affiliation{\RLEaffil}

\author{Ilan~T.~Rosen}
\affiliation{\RLEaffil}

\author{Junyoung~An}
\affiliation{\RLEaffil}
\affiliation{\EECSaffil}

\author{Helin~Zhang}
\affiliation{\RLEaffil}

\author{Aranya~Goswami}
\affiliation{\RLEaffil}

\author{Kate~Azar}
\affiliation{\RLEaffil}
\affiliation{\EECSaffil}
\affiliation{\LLaffil}

\author{Jeffrey~M.~Gertler}
\affiliation{\LLaffil}

\author{Bethany~M.~Niedzielski}
\affiliation{\LLaffil}

\author{Mollie~E.~Schwartz}
\affiliation{\LLaffil}

\author{Terry~P.~Orlando}
\affiliation{\RLEaffil}
\affiliation{\EECSaffil}

\author{Jeffrey~A.~Grover}
\affiliation{\RLEaffil}

\author{Kyle~Serniak}
\affiliation{\RLEaffil}
\affiliation{\LLaffil}

\author{William~D.~Oliver}
\email{william.oliver@mit.edu}
\affiliation{\RLEaffil}
\affiliation{\EECSaffil}
\affiliation{\Physaffil}

\date{\today}

\begin{abstract}
Josephson tunnel junctions are essential elements of superconducting quantum circuits. 
The operability of these circuits presumes a $2\pi$-periodic sinusoidal potential of a tunnel junction, but higher-order corrections to this Josephson potential, often referred to as ``harmonics," cause deviations from the expected circuit behavior.
Two potential sources for these harmonics are the intrinsic current-phase relationship of the Josephson junction and the inductance of the metallic traces connecting the junction to other circuit elements.
Here, we introduce a method to distinguish the origin of the observed harmonics using nearly-symmetric superconducting quantum interference devices (SQUIDs). 
Spectroscopic measurements of level transitions in multiple devices reveal features that cannot be explained by a standard cosine potential, but are accurately reproduced when accounting for a second-harmonic contribution to the model.
The observed scaling of the second harmonic with Josephson-junction size indicates that it is due almost entirely to the trace inductance.
These results inform the design of next-generation superconducting circuits for quantum information processing and the investigation of the supercurrent diode effect.

\end{abstract}
\maketitle

\section{I\MakeLowercase{ntroduction}}
Josephson junctions (JJs) are central to mesoscopic superconductivity, with applications in superconducting quantum circuits \cite{Blais04PRA}, superconducting classical logic \cite{Likharev91IEEE}, metrology \cite{Barone82book}, Josephson diodes \cite{Misaki21PRB}, and superconducting sensors \cite{Peacock96Nature, Barzanjeh20SciAdv}.
A common variant is the tunnel junction, comprising a superconductor-insulator-superconductor (SIS) layer structure.
Within the standard formulation of the Josephson equations, the current-phase relationship of a tunnel junction is  $I(\varphi)=I_c \sin(\varphi)$, where $I_c$ is the critical current and $\varphi$ is the the phase difference across the junction, corresponding to a potential energy proportional to $\cos{(\varphi)}$.
However, provided time-reversal symmetry holds, a more general $2\pi$-periodic potential described by the Fourier components $\cos{(n\varphi)}$ for $n\geq1$ is permitted~\cite{Golubov04RevModPhys}.
These harmonics can significantly impact circuit properties, affecting, for example, qudit-based processors~\cite{Willsch24NatPhy}, coupler architectures~\cite{Putterman25Arxiv, Vanselow25Arxiv}, and observations of the supercurrent diode effect~\cite{Fulton72PRB, Fominov22PRB, Souto22PRL, Bozkurt23Scipost, Paolucci23APL, Nadeem2023NatRevPhy}.

\begin{figure}
    \subfloat{\label{fig1a}}
    \subfloat{\label{fig1b}}
    \includegraphics[]{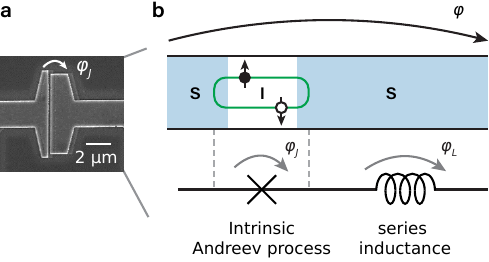}
    \caption{\textbf{Two sources of harmonics in Josephson tunnel junctions.} \textbf{(a)} A scanning electron microscope (SEM) image of an Al/AlOx tunnel junction. \textbf{(b)} Generalized model for a tunnel junction, with $\varphi_J$ the phase across the junction and  $\varphi_L$ the phase across the inductance of the circuit trace leading away from the junction (often neglected). External circuit elements respond to the total phase $\varphi = \varphi_J + \varphi_L$. Harmonic corrections are predicted from the microscopic Andreev model of supercurrent transport as well as from the trace inductance.}
    \label{fig1}
\end{figure}

\begin{figure*}
    \centering
    \subfloat{\label{fig2a}}
    \subfloat{\label{fig2b}}
    \subfloat{\label{fig2c}}
    \subfloat{\label{fig2d}}
    \subfloat{\label{fig2e}}
    \subfloat{\label{fig2f}}
    \includegraphics[]{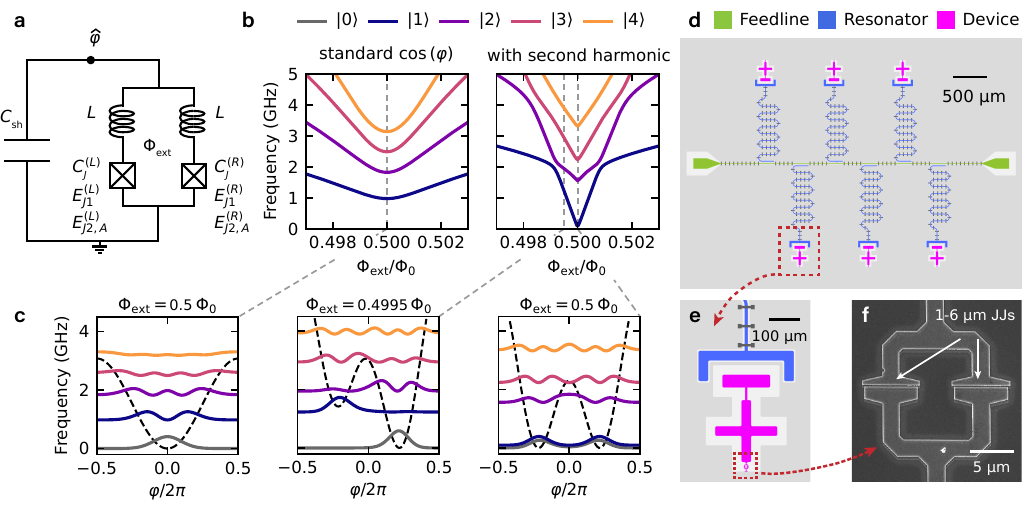}
    \caption{\textbf{Device overview and effects of the second harmonic.}
    \textbf{(a)} Schematic diagram of the device comprising a SQUID and a shunt capacitor $C_{\text{sh}}$. Each arm of the SQUID comprises a series combination of a linear inductance and a Josephson junction. 
    The total phase across the SQUID arms is denoted as $\hat\varphi$.
    The junction is parameterized by the Josephson energies for the fundamental and intrinsic second harmonics, $E_{J1}$ and $E_{J2, A}$, respectively. 
    Each inductor in series with the junction will contribute to the potential similarly to $E_{J2, A}$.
    \textbf{(b)} Expected transition spectra near the external flux-bias $\Phi_{\text{ext}}=0.5\:\Phi_0$ in the absence/presence (left/right) of the second harmonic. Distinct changes in the transition spectra, such as the advent of an avoided crossing and nearly degenerate transition, are expected. 
    \textbf{(c)} The expected potential energy (black dashed lines) and wavefunctions (solid lines) of the energy eigenstates in the absence/presence of the second harmonic (left/middle, right).
    \textbf{(d)} A false-color GDS design of a chip with six SQUIDs for measuring the second-harmonic contribution. 
    \textbf{(e)} Each SQUID is shunted by a nominal capacitance $C_{\text{sh}}=\SI{73}{fF}$ and coupled to its own readout resonator. 
    \textbf{(f)} SEM image of a representative SQUID. The fundamental harmonic $E_{J1}$ is adjusted by varying the length of the junction, $1-\SI{6}{\micro m}$, for a fixed junction width (overlap) of $\SI{0.2}{\micro m}$. The two parallel arms are otherwise nominally identical. Three nominally identical chips were measured in this work.
    }
    \label{fig2}
\end{figure*}

Despite the fact that the harmonic terms ($n\geq2$) have traditionally been neglected, a recent study of Al/AlOx tunnel junctions observed $n=2$ coefficients as large as $10\%$ of the $n=1$ coefficient~\cite{Willsch24NatPhy}, a result further validated by Refs.~\cite{wang2024, Putterman25Arxiv, Vanselow25Arxiv, fechant25arxiv}. 
There are two potential sources of these harmonics: 1) the Andreev processes intrinsic to the JJ and 2) unavoidable inductance arising from the metallic traces connecting the JJ to other circuit elements~[Fig.~\ref{fig1}], but their relative contributions are unknown. 
Because the scaling of the $n\ge2$ coefficients with respect to the $n=1$ coefficient differs depending on the source, to accurately predict the higher-order corrections to the circuit Hamiltonian and the resultant dynamics, it is critical to understand which source dominates. 

In this work, we introduce a method using nearly symmetric SQUIDs to assess the relative contribution and scaling of the harmonic terms with the goal of distinguishing the two sources. 
Flux-biasing a SQUID near half a flux quantum suppresses the fundamental harmonic, increasing the relative prominence of the second-harmonic contribution.
With this method, we observe distinct features in the transition spectra of our devices that cannot be explained by the standard $\cos{(\varphi)}$ potential, but are well-described when the second harmonic is included in the Hamiltonian.
These features are reproducibly observed across multiple devices with varying $n=1$ coefficient. 
Lastly, by comparing the ratios of the $n=2$ to the $n=1$ coefficients in this series of devices, we conclude that the second harmonic in the Josephson potential is primarily due to the series inductance, bounding the intrinsic Andreev harmonics to less than 0.1\% in our devices.

\section{H\MakeLowercase{armonics of a tunnel junction}}
The harmonics of a tunnel junction arise from two possible sources: an Andreev process intrinsic to the junction and a series inductance associated with the circuit trace~[Fig.~\ref{fig1}].
In the presence of both sources, as represented in Fig.~\ref{fig1b}, the generalized potential of this system is
\begin{equation}\label{higher_harmonics}
    U(\varphi) = -E_{J1} \cos{(\varphi)} + E_{J2}\cos{(2\varphi)} - \cdots,
\end{equation}
where $E_{J1} \approx I_c\Phi_0/2\pi$ is the standard Josephson energy, and the effective coefficient $E_{J2}$ for the second harmonic is approximately given by the sum \cite{Suppl}
\begin{equation}\label{effective_EJ2}
    E_{J2} \approx  E_{J2,A} + E_{J2, L}.
\end{equation}
Here, $E_{J2,A}$ and $E_{J2, L}$ are the energies of the second-harmonic contribution due to the intrinsic Andreev process and series inductance, respectively.

\begin{figure*}
    \centering
    \subfloat{\label{fig3a}}
    \subfloat{\label{fig3b}}
    \subfloat{\label{fig3c}}
    \subfloat{\label{fig3d}}
    \subfloat{\label{fig3e}}
    \subfloat{\label{fig3f}}
    \includegraphics[width=1.0\linewidth]{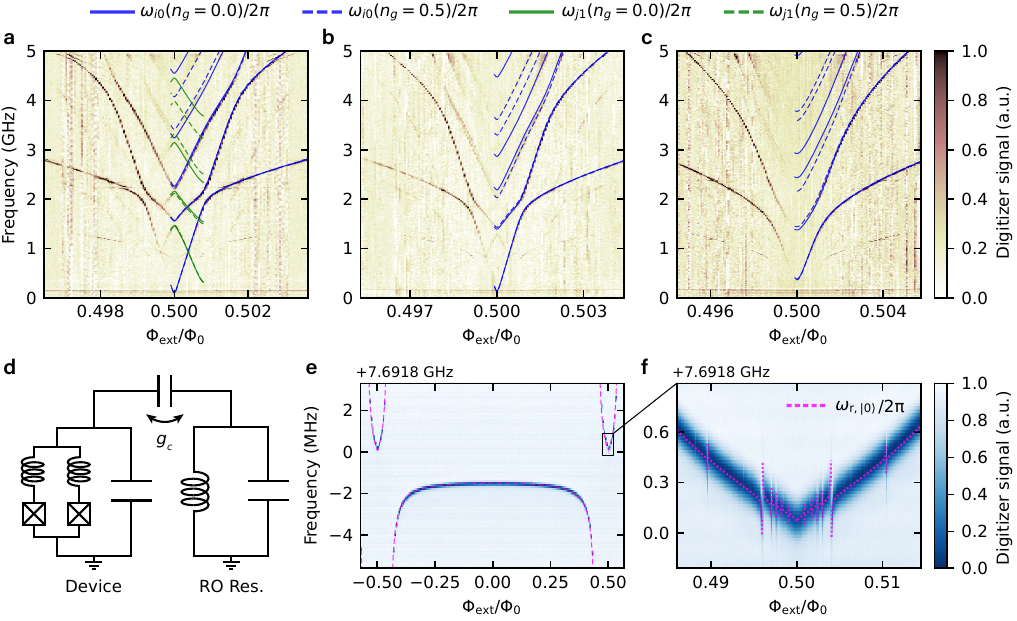}
    \caption{\textbf{Spectroscopic measurements of SQUIDs.} The near-half-flux transition spectra of the \textbf{(a)} $\SI{6}{\micro m}$, \textbf{(b)} $\SI{5}{\micro m}$, and \textbf{(c)} $\SI{4}{\micro m}$ SQUIDs on Chip A, measured by two-tone spectroscopy. The predicted changes due to the second harmonic are clearly observed. The blue solid and dashed lines overlaid on the spectra represent the numerical fits for the transition frequency from the ground state ($\omega_{i0}=\omega_{i}-\omega_0$, $i\geq1$) at gate-charge $n_g=0$ and $n_g=0.5$, respectively. Green lines represent transitions from the first excited states ($\omega_{j1}$, $j\geq2$). Numerical fits are only plotted over part of the measured range for visibility of the spectrum. \textbf{(d)} Schematic diagram of the device capacitively coupled to a readout resonator. \textbf{(e),(f)} Resonator single-tone spectroscopy data of the $\SI{6}{\micro m}$ device [\textbf{(a)}]. The magenta dashed lines indicate the theoretical resonator frequency when the device is in its ground state, $\omega_{r,|0\rangle}=\omega_r+\delta\omega_r$, calculated using circuit parameters obtained from the spectrum fit.}
    \label{fig3}
\end{figure*}

The second-harmonic energies for the two sources scale differently with junction area. 
For constant critical current density, the energy $E_{J1}$ of a junction scales approximately with the overlap area. 
Because the critical current density is fundamentally governed by the intrinsic Andreev process, $E_{J2,A}$ also scales with the area.  
As such, the ratio $E_{J2,A}/E_{J1}$ between the fundamental and intrinsic second harmonic is expected to remain constant as the JJ area is varied.

On the other hand, the second harmonic caused by the series inductance, $E_{J2,L} \approx E_{J1}^2/4E_L$, scales quadratically with $E_{J1}$~\cite{Willsch24NatPhy, Karif17PRA, Putterman25Arxiv, Suppl}, where $E_L = (\Phi_0/2\pi)^2/L$ is the inductive energy of the series inductor.
This formula can be intuitively understood within a semiclassical or mean-field picture as a consequence of current conservation between the junction and the inductor, which locks the inductor phase $\varphi_L$ to the junction phase $\varphi_J$ via the equation $I=\Phi_0\varphi_L/2\pi L\approx I_c \sin{(\varphi_J)}$. 
The energy of the inductor is therefore $LI^2/2\approx LI_c^2\sin^2{(\varphi_J)}/2$, which can be written as $-(E_{J1}^2/4E_L)\cos{(2\varphi_J)}$ if the constant offset is ignored. 
When written in terms of the total phase $\varphi$, this correction term becomes $+(E_{J1}^2/4E_L)\cos{(2\varphi)}$ up to leading order (note the sign flip compared to the correction written in terms of $\varphi_J$).
We stress that while this formula is approximately accurate for the measurements we will present here, it is only valid when the fluctuations of $\varphi_L$ are much smaller than $2\pi$~\cite{Martin23PRX, Egusquiza25PRX, Divincenzo25PRX}.
This is not necessarily true for circuits with delocalized phase states, such as many types of qubits used in quantum computing applications, where the inductance-induced harmonics must be calculated using a full quantum treatment (see Supplemental Material~\cite{Suppl} for discussion). 
For the remainder of this manuscript we will proceed with this approximation $E_{J2,L} \approx E_{J1}^2/4E_L$, which agrees with the full quantum mechanical approach for our devices.

\begin{table*}[!htb]
\caption{\label{table}
\textbf{Characterization of SQUIDs on Chip A.}  The JJ lengths are nominal values, whereas the right-hand side parameters are obtained from the measurement. Circuit parameters were obtained by fitting two-tone device spectroscopy and single-tone resonator spectroscopy data. JJ asymmetry is defined as $(E_{J1}^{(L)}-E_{J1}^{(R)})/(E_{J1}^{(L)}+E_{J1}^{(R)})$.
Apart from the JJ length, all devices were designed with the same nominal parameters: $C_{\text{sh}}=\SI{73}{fF}$, and a JJ width of $\SI{0.2}{\micro m}$. Assuming nominal JJ areas, the average measured critical current density is estimated to be $J_c\approx0.41\:\SI{}{\micro A/\micro m^2}$.}
\begin{ruledtabular}
\begin{tabular}{c|ccccc}
\makecell{JJ length ($\SI{}{\micro m}$)}& 
$E_C/h$ (GHz)&
$E_{J1}^{(L)}/h$ (GHz)&
\makecell{JJ asymmetry} (\%)&
$\omega_r/2\pi$ (GHz)&
$g_c/2\pi$ (MHz) \\
\colrule
2 &0.161  & 84  & 0.74  & 7.5657   & 52.0\\
3 &0.138  & 125 & 1.02  & 7.5891   & 43.6\\
4 &0.118  & 167 & 0.39  & 7.6252   & 38.1\\
5 &0.104  & 203 & 0.27  & 7.6542   & 33.9\\
6 &0.0956 & 231 & 0.33  & 7.6918   & 30.8\\
\end{tabular}
\end{ruledtabular}
\end{table*}

For both of the aforementioned sources, the second harmonic is expected to be a few orders of magnitude smaller than the fundamental harmonic.
Therefore, to robustly resolve the effect of the second harmonic on circuit behavior, we leverage the suppression of the fundamental harmonics provided by a DC SQUID \cite{Larsen20PRL, Smith20npjQI}. 
We connect two nominally identical junctions in parallel to form a SQUID, which is then shunted by a capacitor [Fig.~\ref{fig2a}].
This capacitively shunted SQUID is biased by an external flux close to half a flux quantum, which we refer to as ``half flux.''
Note that our devices are similar to conventional flux-tunable transmons \cite{Koch07PRA}, but feature much higher $E_{J1}$ as we will present later.
Using the expressions for the fundamental and second harmonics in Eqs.~\eqref{higher_harmonics}--\eqref{effective_EJ2}, the potential energy of the SQUID can be approximated as
\begin{align}
    &U_\text{SQUID}(\phi_{\text{ext}}) \approx\\ \nonumber
    &\:\:\:\:\:\:-E_{J1}^{(L)}\cos{(\hat\varphi)}+E_{J2}^{(L)}\cos{(2\hat\varphi)}\\ \nonumber
    &\:\:\:\:\:\: -E_{J1}^{(R)}\cos{(\hat\varphi-\phi_{\text{ext}})}+E_{J2}^{(R)}\cos{\big(2(\hat\varphi-\phi_{\text{ext}})\big)},
\end{align}
where $E_{Jn}^{(L\, (R)\, )}$ is the energy of the $n$th harmonic for the left (right) arm of the SQUID, and $\phi_{\text{ext}}=2\pi\Phi_{\text{ext}}/\Phi_0$ is the reduced external flux.
To calculate the transition spectrum of the device, we here and henceforth promote the phase variable to a quantum operator.
When biased at exactly half flux, the potential becomes 
\begin{align}\label{potential_at_half}
    U_\text{SQUID}(\pi) &\approx 
   -\left(\Delta E_{J1}\right)\cos{(\hat\varphi)} + \left(\Sigma E_{J2}\right) \cos{(2\hat\varphi)},
\end{align}
where $\Delta E_{J1} = E_{J1}^{(L)}-E_{J1}^{(R)}$, is the difference in the fundamental harmonic of the left and right SQUID arms, and $\Sigma E_{J2}=E_{J2}^{(L)}+E_{J2}^{(R)}$ is the sum of the second harmonic of the two arms.  
Near half-flux bias in the SQUID loop, the fundamental harmonic of the SQUID ($2e$ charge transport) is suppressed, while the second harmonic ($4e$ charge transport) is enhanced.
This suppression and enhancement is analogous to that observed with the Aharonov-Bohm effect.

Using this second-harmonic approximation, the Hamiltonian of the capacitively shunted SQUID is then given by \cite{Suppl}
\begin{equation}\label{squid_hamiltonian}
    H_{\text{SQUID}}(\phi_\text{ext}) = 4E_C (\hat{n}-n_g)^2 + U_{\text{SQUID}}(\phi_\text{ext}),
\end{equation}
where the charging energy $E_C = e^2/2\big(C+C_J^{(L)}+C_J^{(R)}\big)$ includes the shunt capacitance $C$ and the left and right junction capacitances $C_J^{(L)}$ and $C_J^{(R)}$, $\hat{n}$ is the Cooper-pair number operator, and $n_g$ denotes the gate charge of the SQUID.
Figure~\ref{fig2b} shows the transition spectra of a representative device near half flux, in the absence and presence of the second harmonic, obtained by diagonalizing Eq.~\eqref{squid_hamiltonian}. 
The second harmonic significantly alters the spectrum and wavefunctions. 
Near half flux, the presence of the second harmonic results in the formation of a double-well potential [Fig.~\ref{fig2c}]. 
The device transition spectrum then consists of inter-well transitions with characteristic V-shaped flux dispersion as well as intra-well transitions with weaker flux dispersion. 
At exactly half flux, the minima of the two wells are equal in energy, resulting in de-localized nearly-degenerate ground states. 
This structure is reminiscent of persistent-current-based qubits such as the flux qubit~\cite{Orlando99PRB},  fluxonium~\cite{Manu09Science}, and the kite qubit~\cite{Smith20npjQI}.

Figure~\ref{fig2d} shows the design of the experimental chip containing six capacitively shunted SQUIDs.
Each device is capacitively coupled to its own readout resonator [Fig.~\ref{fig2e}] with resonance frequencies separated by about 30 MHz from one another. 
All resonators are inductively coupled to a shared feedline.
As shown in Fig.~\ref{fig2f}, the Dolan-style JJs are fabricated from shadow-evaporated Al with an AlOx insulating barrier.
The fundamental harmonic $E_{J1}$ of each SQUID was adjusted by changing the JJ length from $\SI{1}{\micro m}$ to $\SI{6}{\micro m}$. 
Apart from the JJ length, the remainder of the device parameters were nominally identical.
Although six devices were designed on a single chip, we only measured five because the charge dispersion of the $\SI{1}{\micro m}$ device was too large near half flux.
The measured fundamental harmonic $E_{J1}/h$ of the five SQUIDs ranged approximately from $\SI{80}{GHz}$ to $\SI{230}{GHz}$.
Note that these values are much larger than typical $E_{J1}/h$ for a transmon, which is on the order of $\SI{10}{GHz}$ \cite{Houck07Nature}.
Further parameter values are listed in Table~\ref{table}.
Three nominally identical chips $-$ each containing six SQUIDs and labeled Chip~A, Chip~B and Chip~C $-$ were fabricated on the same wafer in 2020 and tested from Aug. 2024 (Chip~B) to Feb. 2025 (Chip~A, Chip~C).

\section{S\MakeLowercase{pectroscopic evidence for the second harmonic}}
To probe the second harmonic, we perform two-tone spectroscopy near half flux. 
Here, the external flux is adjusted using a superconducting coil mounted on the lid of the device package. 
We send two microwave tones through the feedline~[Fig.~\ref{fig2d}]: a drive tone to excite the circuit and a readout tone to detect the energy-level population via frequency shift of the readout resonators. 
For each external flux value, the drive tone frequency is swept, while the readout tone is fixed at the resonator frequency.

Figures~\ref{fig3a}-\ref{fig3c} show the measured transition spectra for the SQUIDs with JJ lengths of $\SI{6}{\micro m}$, $\SI{5}{\micro m}$, and $\SI{4}{\micro m}$, respectively, on Chip~A. 
The observed results cannot be explained by the conventional cosine potential [Fig.~\ref{fig2b}]. 
We fit the measured spectra by numerically diagonalizing the Hamiltonian in Eq.~\eqref{squid_hamiltonian}, and extract the circuit parameters $E_C$, $E_{J1}^{(L)}$, $\Delta E_{J1}$, and $E_{J2}^{(L)}$ (See Supplemental Material \cite{Suppl} for details). 
Due to the suppression of $E_{J1}$ near half flux, the effective ratio between the Josephson and capacitive energies is reduced, leading to an increased charge dispersion for higher transitions; this effect is well-captured by the second harmonic model. 
Finally, in Fig.~\ref{fig3a}, transitions between the first excited state and higher states are visible due to thermal population, while faint features near $\SI{1}{GHz}$ correspond to multi-photon transitions (see Supplemental Material~\cite{Suppl} for details). 
Overall, we observe excellent agreement between the measured data and the numerical fits.

In addition to measuring the transition spectra, we perform single-tone spectroscopy of the readout resonators by sending a variable-frequency readout tone through the device feedline.
Due to the flux-dependent device Hamiltonian, the resonator experiences a flux-dependent dispersive shift.
Figures~\ref{fig3e}~and~\ref{fig3f} show the measured resonator spectroscopy of the $\SI{6}{\micro m}$ device on Chip A. 
Modeling the capacitive interaction between the device and resonator as $\hbar g_c \hat{n} (\hat{a} + \hat{a}^\dagger)$ [Fig.~\ref{fig3d}], where $\hat{a}$ is the lowering operator for resonator excitations, and using the device parameters from the transition spectrum fit [Fig.~\ref{fig3a}], we calculate the dispersive shift as 
\begin{equation}
    \delta\omega_r = |g_c|^2\sum_{i\neq 0} |\langle i|\hat{n}|0\rangle|^2 \frac{-2\omega_{i0}}{\omega_{i0}^2-\omega_r^2},
\end{equation}
where $|i\rangle$ and $\omega_{i0}/2\pi$ are the $i$th bare energy eigenstate of the device and transition frequency from its ground state, respectively. The bare resonator frequency $\omega_r/2\pi$ is measured with a high-power readout tone, and $g_c$ is the coupling strength between the device and the resonator.  
The only additional free parameter is the coupling strength, which we find to be $g_c/2\pi\approx \SI{31}{MHz}$. The coupling strengths for other four devices are listed in Table.~\ref{table}.
The excellent agreements of both the transition spectra and dispersive shift with theory indicate that the Hamiltonian Eq.~\eqref{squid_hamiltonian} accurately captures the device properties.

\begin{figure}
    \includegraphics[]{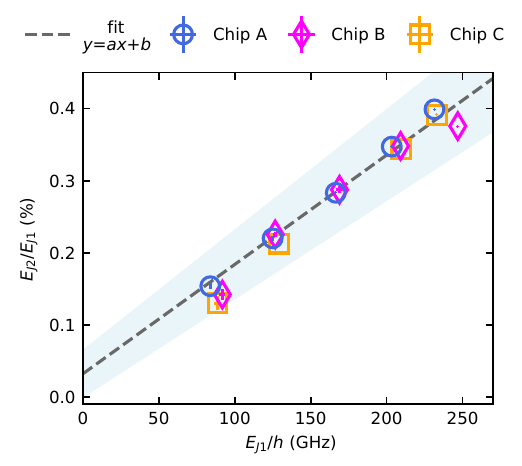}
    \caption{\textbf{Measured $E_{J2}/E_{J1}$ as a function of $E_{J1}$.} The ratio and $E_{J1}$ were extracted from the spectra fits. The horizontal and vertical error bars centered on each marker represent the spectrum fit errors for $E_{J1}$ and $E_{J2}/E_{J1}$, respectively. The gray dashed linear fit is obtained by performing Orthogonal Distance Regression (ODR) over all data points with both fit errors as input uncertainties. The shaded region represents the uncertainty in the ODR fit, determined from the standard deviations of the estimated slope and intercept. A linear trend is observed in all three chips. } 
    \label{fig4}
\end{figure}

\section{D\MakeLowercase{etermining the source of harmonics}}

Equipped with the extracted $E_{J1}, E_{J2}$ from the spectra fits, we can estimate the relative contributions of the intrinsic Andreev process and the series inductance to the second harmonic.
Due to the linear and quadratic scaling of $E_{J2, A}$ and $E_{J2, L}$ with junction area, respectively, the ratio
\begin{equation}
    \frac{E_{J2}}{E_{J1}} \approx \beta + \left(\frac{1}{4E_L}\right)E_{J1}
\end{equation}
is expected to be a linear function of $E_{J1}$, where the first term $\beta=E_{J2, A}/E_{J1}$ (constant in $E_{J1}$) represents the contribution from the intrinsic Andreev process, and the second term the contribution from the series inductance. 
From the spectra fits, we extract $E_{J2}/E_{J1}$ across five SQUIDs with varying $E_{J1}$. 
This process is repeated for the three nominally identical chips, and the results are shown in Fig.~\ref{fig4}.
Note that the $\SI{4}{\micro m}$ device of Chip C (orange square) is excluded because fabrication imperfections resulted in a large JJ asymmetry of approximately 10$\%$, masking signatures of the second harmonic.

We observe the predicted linear dependence of the ratio $E_{J2}/E_{J1}$ on $E_{J1}$ and extract the $y$-intercept and slope using an Orthogonal Distance Regression (ODR).
From the intercept, we bound the contribution of the JJ-intrinsic Andreev processes $E_{J2,A}$ to be less than 0.1\% of $E_{J1}$, roughly one to two orders of magnitude smaller than the values reported in Ref.~\cite{Willsch24NatPhy} and similar to the result in Ref.~\cite{Vanselow25Arxiv}.
Within the duration of the measurements, we also did not observe noticeable aging effects~\cite{Willsch24NatPhy}.
From the slope of the ODR fit, we extract a series inductance of $L_{\text{fit}}=10\pm1\:\SI{}{pH}$. 
Finally, to check this observation, we numerically simulate the inductance of the SQUID loop using the superconducting circuit simulation software \nobreak{\textit{InductEx}} \cite{Fourie05IEEE}. 
The simulated inductances for the five SQUID loops ranged from $L_{\text{sim}}=\SI{10.2}{pH}$ to $\SI{10.7}{pH}$ due to the the varying JJ length ($\SI{6}{\micro m}-\SI{2}{\micro m}$). These simulated values are in agreement with the measured inductance within uncertainty. We therefore conclude that the observed second harmonic is primarily due to series inductance.

\section{D\MakeLowercase{iscussion}}
We have demonstrated a method to measure junction harmonics and estimate the relative contributions of the Andreev process and series inductance. 
Using this method, we reproducibly observe distinct changes in transition spectra that cannot be explained by the standard cosine potential but are accurately captured by including the second-harmonic term in the potential energy of the circuit.

Although it was suggested that the mesoscopic model of the AlOx barrier intrinsic to tunnel junctions could predict percent-level harmonics~\cite{Willsch24NatPhy}, in our experiment using standard junction fabrication the observed trend of the $E_{J2}/E_{J1}$ ratio as a function of $E_{J1}$ is linear and indicates that the series inductance is almost entirely responsible for the second harmonic.
We note that the intrinsic harmonics may be influenced by underlying details of the fabrication process or junction aging, which could be the subject of a future experiment.
Furthermore, our results provide experimental verification of claims in Refs.~\cite{Putterman25Arxiv, Vanselow25Arxiv}, which attribute harmonics observed in microwave-activated couplers to series inductance.

These results highlight the need to account for the dressing effects of surrounding circuitry when attempting to probe the physics of mesoscopic devices. 
As a closely related example, consider what we would expect if the low-frequency response of our SQUIDs were measured using transport techniques. 
Away from half flux, broken time-reversal symmetry would result in a supercurrent diode effect, in which the magnitude of the SQUID critical current would be different depending on the sign of the current (see Supplemental Material for further discussion). 
Again, we would find that, although our junctions are standard tunnel junctions, the system would demonstrate ``anomalous'' behavior due to the presence of the trace inductance~\cite{Fulton72PRB,Haenel22arxiv, Paolucci23APL, Nadeem2023NatRevPhy}.

Accurately capturing the origin of observed harmonics is thus essential for correctly predicting the dynamics of complex circuits containing flux-biased high-$E_J$ junctions, such as the quarton~\cite{Ye24SciAdv}, SNAIL~\cite{Chapman23PRXQ}, asymmetrically threaded SQUIDs~\cite{Putterman25Arxiv, Vanselow25Arxiv} and arbitrary supercurrent diodes~\cite{Bozkurt23Scipost}.
As such, we expect the harmonic characterization methods presented here to guide the design and development of novel superconducting circuits and probes of novel Josephson junction physics.

\textbf{Acknowledgments}
We gratefully acknowledge fruitful discussions with Ioan Pop and Manuel Houzet.
This research is sponsored in part by the U.S. Army Research Office Grant No. W911NFF-23-1-0045 (Extensible and Modular Advanced Qubits), in part by the U.S. Department of Energy, Office of Science, National Quantum Information Science Research Centers, Co-design Center for Quantum Advantage (C2QA) under contract number DE-SC0012704, and in part under Air Force Contract No. FA8702-15-D-0001.
J.K. and J.A. gratefully acknowledge support from the Korea Foundation for Advanced Studies (KFAS).
M.H. is supported by an appointment to the Intelligence Community Postdoctoral Research Fellowship Program at the Massachusetts Institute of Technology administered by Oak Ridge Institute for Science and Education (ORISE) through an interagency agreement between the U.S. Department of Energy and the Office of the Director of National Intelligence (ODNI).
The views and conclusions contained herein are those of the authors and should not be interpreted as necessarily representing the official policies or endorsements, either expressed or implied, of the U.S. Air Force or the U.S. Government.

\bibliography{refs}

\end{document}


\title{Supplemental Material for\\``Emergent Harmonics in Josephson Tunnel Junctions Due to Series Inductance''}

\def\RLEaffil{Research Laboratory of Electronics, Massachusetts Institute of Technology, Cambridge, MA 02139, USA}
\def\LLaffil{Lincoln Laboratory, Massachusetts Institute of Technology, Lexington, MA 02421, USA}
\def\Physaffil{Department of Physics, Massachusetts Institute of Technology, Cambridge, MA 02139, USA}
\def\EECSaffil{Department of Electrical Engineering and Computer Science, Massachusetts Institute of Technology, Cambridge, MA 02139, USA}

\author{Junghyun~Kim}
\email{kimjung@mit.edu}
\affiliation{\RLEaffil}
\affiliation{\EECSaffil}

\author{Max~Hays}
\email{maxhays@mit.edu}
\affiliation{\RLEaffil}

\author{Ilan~T.~Rosen}
\affiliation{\RLEaffil}

\author{Junyoung~An}
\affiliation{\RLEaffil}
\affiliation{\EECSaffil}

\author{Helin~Zhang}
\affiliation{\RLEaffil}

\author{Aranya~Goswami}
\affiliation{\RLEaffil}

\author{Kate~Azar}
\affiliation{\RLEaffil}
\affiliation{\EECSaffil}
\affiliation{\LLaffil}

\author{Jeffrey~M.~Gertler}
\affiliation{\LLaffil}

\author{Bethany~M.~Niedzielski}
\affiliation{\LLaffil}

\author{Mollie~E.~Schwartz}
\affiliation{\LLaffil}

\author{Terry~P.~Orlando}
\affiliation{\RLEaffil}
\affiliation{\EECSaffil}

\author{Jeffrey~A.~Grover}
\affiliation{\RLEaffil}

\author{Kyle~Serniak}
\affiliation{\RLEaffil}
\affiliation{\LLaffil}

\author{William~D.~Oliver}
\email{william.oliver@mit.edu}
\affiliation{\RLEaffil}
\affiliation{\EECSaffil}
\affiliation{\Physaffil}

\maketitle

\tableofcontents

\clearpage
\section{Experimental setup}
\begin{figure}[!htb]
    \centering
    \includegraphics[]{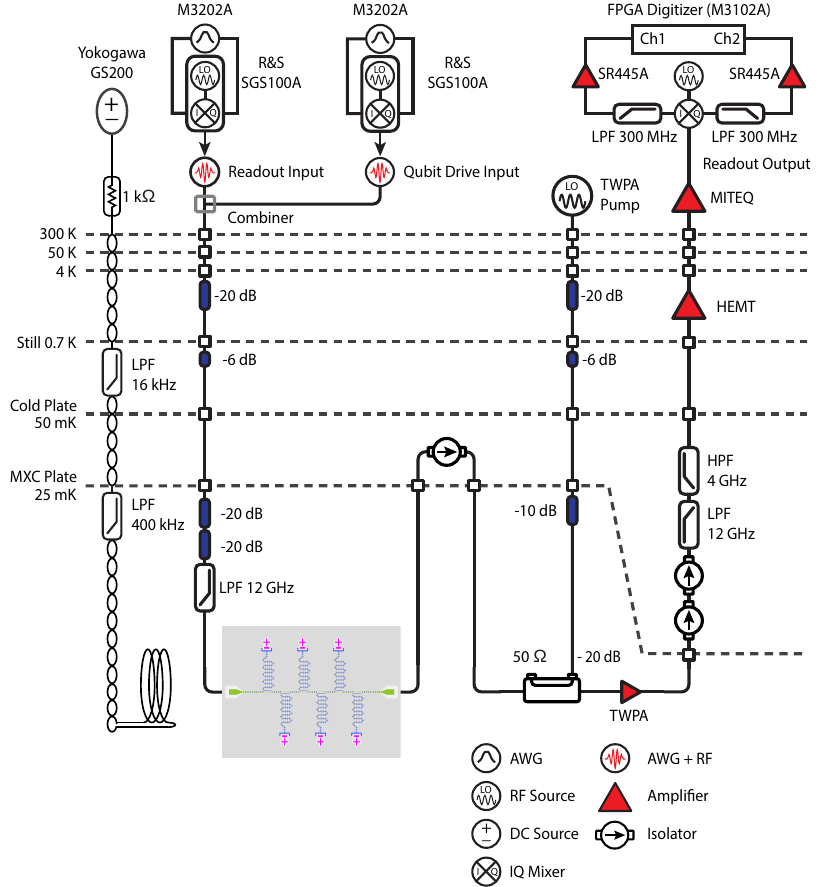}
    \caption{Wiring diagram of the experimental setup.}
    \label{wiring}
\end{figure}

Figure~\ref{wiring} represents the experimental wiring.
All experiments were conducted in a Leiden CF-450 dilution refrigerator operating at a base temperature of approximately 25 mK. 
The flux bias for the five SQUIDs is provided by a superconducting coil mounted to the lid of the sample package. 
The equipment used in this experiment is given in Table~\ref{equipment}. 

\begin{table}[!htb]
\caption{\label{equipment}
\textbf{Summary of control equipment.}
}
\begin{tabular}{|c|c|c|}
\hline
Component & Manufacturer & Model\\
\hline
Dilution Refrigerator & Leiden Cryogenics & CF-450\\
RF Source & Rohde \& Schwarz & SGS100A\\
DC Source & Yokogawa & GS200\\
Control Chassis & Keysight & M9019A\\
\makecell{Arbitrary Waveform Generator (AWG)} & Keysight & M3202A\\
Digitizer & Keysight & M3102A\\
\hline
\end{tabular}
\end{table}

\newpage
\section{Potential in the presence of Andreev harmonics and series inductance}
\subsection{Andreev process}
\noindent The potential energy of a Josephson tunnel junction described by the Andreev process is given by 
\begin{equation}\label{andreev_harmonics}
\begin{aligned}
    U(\varphi_J) &= -\Delta\sum_{n=1}^{N}\sqrt{1-T_n\sin^2\left(\frac{\varphi_J}{2}\right)},
\end{aligned}
\end{equation}
where $T_n$ is the transparency of the $n^{\text{th}}$ conduction channel, and $N$ is the number of the channels. $\varphi_J$ is the phase across the junction.
For a superconductor-insulator-superconductor (SIS) tunnel junction, these two intrinsic parameters are estimated to be $N\sim10^6$, and $T_n \sim 10^{-6}$ \cite{Willsch24NatPhy}. Assuming $T_n$ is small, we can expand the square root to obtain
\begin{equation}
    U(\varphi_J) = -\Delta\sum_{n=1}^{N} \left( {1-\frac{1}{2}T_n\sin^2{\left(\frac{\varphi_J}{2}\right)}-\frac{1}{8}T_n^2\sin^4{\left(\frac{\varphi_J}{2}\right)}-\frac{1}{16}T_n^3\sin^6{\left(\frac{\varphi_J}{2}\right)}-\cdots} \right).
\end{equation}
Using trigonometric identities, the potential can be expressed as a sum of harmonics
\begin{equation}
    U(\varphi_J) = -E_{J1, A} \cos{(\varphi_J)} + E_{J2, A} \cos{(2\varphi_J)} - E_{J3, A} \cos{(3\varphi_J)} + \cdots,
\end{equation}
where the coefficients satisfy
\begin{align}
    E_{J1, A} &= \Delta \sum_{n=1}^{N} \left(\frac{1}{4}T_n + \frac{1}{16}T_n^2 + \frac{15}{512}T_n^3+\cdots\right)\\
    E_{J2,A} &= \Delta \sum_{n=1}^{N} \left(\frac{1}{64}T_n^2 + \frac{3}{256}T_n^3 + \frac{35}{4096}T_n^4-\cdots\right)\\
    E_{J3,A} &= \Delta \sum_{n=1}^{N} \left(\frac{1}{512}T_n^3+\frac{5}{2048}T_n^4+\frac{315}{131072}T_n^5+\cdots \right).
\end{align}

\noindent Taking only the leading order term for each $E_{J1}$ and $E_{J2}$, the coefficients for the intrinsic Andreev process can be approximated as
\begin{align}
    E_{J1,A} &\approx \Delta \sum_{n=1}^{N} \frac{1}{4}T_n\\
    E_{J2,A} &\approx \Delta \sum_{n=1}^{N} \frac{1}{64}T_n^2
\end{align}
Assuming a large number of channels $N\gg1$ and that the transparency follows a distribution $T\sim\rho(T)$, the summation over all conduction channels can be approximated as an integral over the distribution \cite{Willsch24NatPhy}.
\begin{equation}
    \sum_{n=1}^{N} f(T_n) \approx N\int_{0}^{1} \rho(T)f(T)dT.
\end{equation}
The ratio between the first and the second harmonics becomes
\begin{equation}\label{andreev_harmonics_ratio}
    \frac{E_{J2,A}}{E_{J1,A}} = \frac{\frac{1}{64}\sum_{n=1}^{N} T_n^2}{\frac{1}{4}\sum_{n=1}^{N} T_n} \approx \frac{\frac{N}{64}\int_{0}^{1}T^2\rho(T)dT}{\frac{N}{4}\int_{0}^{1}T\rho(T)dT}=\frac{1}{16}\frac{\int_{0}^{1}T^2\rho(T)dT}{\int_{0}^{1}T\rho(T)dT},
\end{equation}
which only depends on the distribution of the transparency.
In this study, we assume our junction size is large enough to provide a sufficiently large number of channels, so that the summation of the transparency (squared) is well approximated by the integral. Moreover, we assume the same transparency distribution $\rho(T)$ for all devices measured, so that varying the junction area only affects the number of channels $N$, which does not appear in the final ratio. 
This assumption is reasonable for our devices because they were fabricated in the same deposition/oxidation run. 
Under these assumptions, the ratio of the second harmonic
\begin{equation}
    \frac{E_{J2,A}}{E_{J1,A}} \approx \frac{1}{16}\frac{\int_{0}^{1}T^2\rho(T)dT}{\int_{0}^{1}T\rho(T)dT} = \beta
\end{equation}
remains constant across the devices under test.  

\subsection{Full-circuit model vs second-harmonic approximation}
In this section, we discuss the validity of the harmonic approximation used to model the effect of the inductance in series with the junction.
We compare the harmonic approximation of the SQUID potentials to a full-circuit analysis, and compute the difference between the energy eigenvalues of the two models for the validity of the harmonic approximation. 
We find that while the harmonic approximation is valid for our circuit parameters, it is not necessarily valid for more typical transmon parameters and therefore should be used with caution.

\begin{figure}[h]
    \centering
    \includegraphics[]{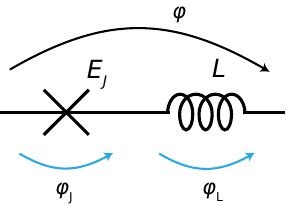}
    \caption{A Josephson junction in series with a linear inductor. The junction capacitance is temporarily neglected.}
    \label{junction_inductor}
\end{figure}

We first derive the harmonic approximation used in the main text, following Ref.~\cite{Willsch24NatPhy}.
Figure~\ref{junction_inductor} shows a Josephson junction in series with an inductor, where the junction capacitance has temporarily been neglected. In addition, in this section we will assume that there are no intrinsic higher harmonics.

\noindent The branch phase variables for the junction and the inductor are denoted as $\varphi_J$ and $\varphi_L$, respectively, and $\varphi = \varphi_J+\varphi_L$ represents the total phase across the series. 
The potential of this circuit is given by 
\begin{equation}\label{series_potential}
    U(\varphi, \varphi_J) = -E_J\cos{(\varphi_J)}+\frac{1}{2L}\left(\frac{\Phi_0}{2\pi}\right)^2(\varphi-\varphi_J)^2.
\end{equation}
To eliminate $\varphi_J$ from the potential, we make the approximation that the current through the junction is equal to the current through the inductor: 
\begin{equation}
    I_c\sin{\varphi_J}=\frac{1}{L}\left(\frac{\Phi_0}{2\pi}\right)(\varphi-\varphi_J),
\end{equation}
which is equivalent to
\begin{equation}\label{current_conservation}
    E_J\sin{\varphi_J}=E_L(\varphi-\varphi_J),
\end{equation}
where $E_L = \frac{1}{L}\left(\frac{\Phi_0}{2\pi}\right)^2$ is the inductive energy.
As discussed in more detail below, we stress that this approximation neglects the junction capacitance and is therefore only valid for certain circuit parameter regimes and should be carefully checked. 
Under this approximation, the $\varphi_J$ degree of freedom can be eliminated from the potential by substituting Eq.~\eqref{current_conservation} into Eq.~\eqref{series_potential}.

To this end, we first present a simple method of successive approximation to calculate the leading-order second-harmonic coefficient. Rewriting Eq.~\eqref{current_conservation}, we obtain
\begin{equation}
    \varphi-\varphi_J = \frac{E_J}{E_L} \sin{\varphi_J}.
\end{equation}
Assuming $E_J\ll E_L$, the zeroth-order equation for $\varphi_J$ is given by
\begin{equation}
    \varphi-\varphi_J^{(0)} = 0.
\end{equation}
Consequently, we find the first-order equation as
\begin{equation}
    \varphi-\varphi_J^{(1)} = \frac{E_J}{E_L}\sin{\varphi_J^{(0)}}=\frac{E_J}{E_L}\sin{\varphi}.
\end{equation}
Substituting $\varphi_J^{(1)}$ into Eq.~\eqref{series_potential} yields
\begin{equation}
\begin{aligned}
    U(\varphi) &= -E_J\cos\left(\varphi -\frac{E_J}{E_L}\sin{\varphi}\right)+\frac{1}{2}E_L\left(\frac{E_J}{E_L}\sin{\varphi}\right)^2\\
    &\approx-E_J\left[ \cos\varphi +\frac{E_J}{E_L}\sin^2{\varphi}\right]+ \frac{1}{2}E_L\left(\frac{E_J}{E_L}\sin{\varphi}\right)^2\\
    &= -E_J\cos\varphi-\frac{1}{2}\frac{E_J^2}{E_L}\sin^2{\varphi}
\end{aligned}
\end{equation}
Using the double-angle formula and dropping the constant term, we obtain
\begin{equation}\label{ind_harmonics_second_0}
\begin{aligned}
    U(\varphi) \approx -E_J\cos{\varphi} + \frac{E_J^2}{4E_L}\cos{(2\varphi)}.
\end{aligned}
\end{equation}

\noindent We also derive the above result following the more detailed method discussed in Ref.~\cite{Willsch24NatPhy}. The generalized harmonic expansion is given by
\begin{equation}\label{ind_harmonics}
    U(\varphi) = -\sum_{n\geq1} E_{Jn, L} \cos{(n\varphi)},
\end{equation}
where the coefficients $E_{Jn,L}$ satisfy 
\begin{equation}\label{fourth_ind_harmonics}
\begin{aligned}
    E_{J1,L} &= E_J \left[1-\frac{1}{8} \left(\frac{E_J}{E_L}\right)^2+\frac{1}{192}\left(\frac{E_J}{E_L}\right)^4+\cdots\right]\\
    E_{J2,L} &= E_J \left[-\frac{1}{4} \left(\frac{E_J}{E_L}\right)+\frac{1}{12}\left(\frac{E_J}{E_L}\right)^3-\frac{1}{96}\left(\frac{E_J}{E_L}\right)^5+\cdots\right]\\
    E_{J3,L} &= E_J\left[\frac{1}{8}\left(\frac{E_J}{E_L}\right)^2-\frac{9}{128}\left(\frac{E_J}{E_L}\right)^4+\frac{1}{64}\left(\frac{E_J}{E_L}\right)^6+\cdots\right]\\
    E_{J4,L} &= E_J\left[-\frac{1}{12}\left(\frac{E_J}{E_L}\right)^3+\frac{1}{15}\left(\frac{E_J}{E_L}\right)^5-\frac{101}{4608}\left(\frac{E_J}{E_L}\right)^7 + \cdots\right].
\end{aligned}
\end{equation}
Truncating the summation at the second harmonic $(n=2)$ and retaining only the leading order terms, the potential is approximated as
\begin{equation}\label{ind_harmonics_second}
    U(\varphi) \approx -E_{J}\cos{(\varphi)} + \frac{E_J^2}{4E_L}\cos{(2\varphi)},
\end{equation}
which is the same as Eq.~\eqref{ind_harmonics_second_0}.
By applying Eq.~\eqref{ind_harmonics_second} to each arm of the SQUID, which comprises a series connection of a junction and an inductor, the Hamiltonian of the circuit is approximately given by

\begin{equation}\label{simplified_Hamiltonian}
\begin{aligned}
    H_{\text{SQUID}}(\phi_\text{ext}) = 4E_C (\hat{n}-n_g)^2 - E_{J}^{(L)}\cos{(\hat\varphi)} -E_{J}^{(R)}\cos{(\hat\varphi-\phi_{\text{ext}})}\\
    +\frac{\left(E_{J}^{(L)}\right)^2}{4E_L}\cos{(2\hat\varphi)}+\frac{\left(E_{J}^{(R)}\right)^2}{4E_L}\cos{\big(2(\hat\varphi-\phi_{\text{ext}})\big)},
\end{aligned}
\end{equation}

\noindent where $E_C=e^2/2(C_J^{(L)}+C_J^{(R)}+C_{\text{sh}})$ includes the shunt and junction capacitances.

\begin{figure}[h]
    \centering
    \includegraphics[]{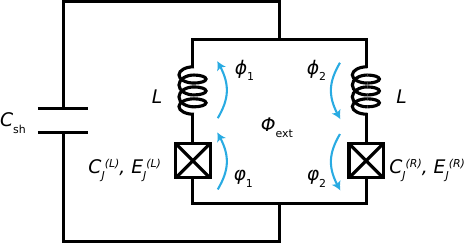}
    \caption{A full-circuit model of the capacitively shunted SQUID including two series inductors.}
    \label{full_circuit}
\end{figure}

\begin{figure}[h] 
    \centering
    \subfloat{\label{full_circuit_spectra_a}}
    \subfloat{\label{full_circuit_spectra_b}}
    \subfloat{\label{full_circuit_spectra_c}}
    \subfloat{\label{full_circuit_spectra_d}}
    \includegraphics[]{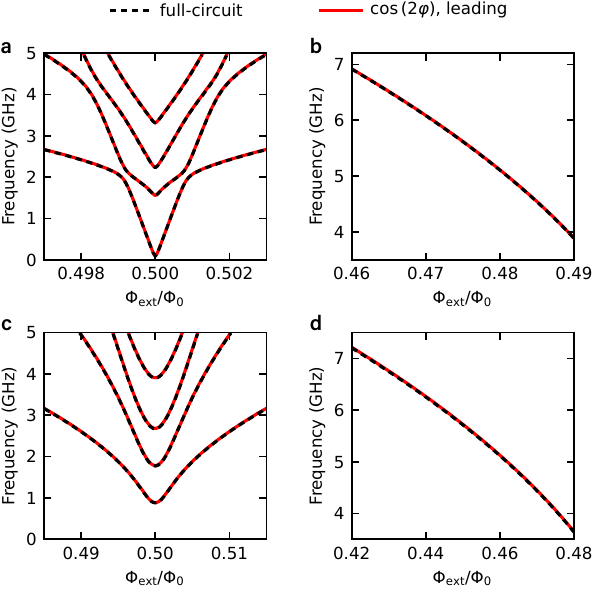}
    \caption{Transition spectra of the full-circuit model (black dashed) and second-harmonic model (red solid). 
    \textbf{(a)} and \textbf{(b)} show the spectra of the $\SI{6}{\micro m}$ device with $E_J\approx \SI{230}{GHz}$, and \textbf{(c)} and \textbf{(d)} of the $\SI{2}{\micro m}$ device with $E_J\approx \SI{80}{GHz}$.}
    \label{full_circuit_spectra}
\end{figure}

We now proceed with a full-circuit analysis to check the validity of this second-harmonic model. 
While above we set the junction current equal to the inductor current in order to eliminate the junction phases from the potential energy, here we will keep these phases and solve the Hamiltonian accounting for all three degrees of freedom~[Fig.~\ref{full_circuit}]. 
The full-circuit Hamiltonian is derived from the following Lagrangian
\begin{equation}
\begin{aligned}
    T &= \frac{1}{2}C_J^{(L)}\left(\frac{\Phi_0}{2\pi}\right)^2\dot\varphi_1^2+\frac{1}{2}C_J^{(R)}\left(\frac{\Phi_0}{2\pi}\right)^2\dot\varphi_2^2+\frac{1}{2}C_{\text{sh}}\left(\frac{\Phi_0}{2\pi}\right)^2(\dot\phi_1+\dot\varphi_1)^2\\
    U &= \frac{1}{2}E_L(\phi_1^2+\phi_2^2)-E_J^{(L)}\cos{(\varphi_1)}-E_J^{(R)}\cos{(\varphi_2)}\\
    L_{\text{full}} &= T-U.
\end{aligned}
\end{equation}
By changing the variables as follows~\cite{Smith20npjQI}
\begin{equation}
\begin{aligned}
\varphi_1 &= \frac{\phi}{2}+\varphi\\
\varphi_2 &= \frac{\phi}{2}-\varphi\\
\phi_1 &= \theta - \frac{1}{2}\phi+\frac{\phi_{\text{ext}}}{2}\\
\phi_2 &= -\theta - \frac{1}{2}\phi+\frac{\phi_{\text{ext}}}{2},
\end{aligned}
\end{equation}
we reach the full-circuit Hamiltonian
\begin{equation}\label{full_circuit_Hamiltonian}
\begin{aligned}
    H_{\text{full}} &= 4E_{C\theta} \:\hat n_\theta^2+ 4E_{C\phi} \:\hat n_\phi^2+ 4E_{C\varphi} \:(\hat n_\varphi-n_g)^2-J\hat n_\theta (\hat n_\varphi-n_g) \\
    &\:\:\:\:\:\:\:\:\:+ E_L\hat \theta^2+\frac{1}{4}E_L(\hat \phi-\phi_{\text{ext}})^2-E_J^{(R)}\cos{\left(\frac{\hat\phi}{2}-\hat\varphi\right)}-E_J^{(L)}\cos{\left(\frac{\hat \phi}{2}+\hat\varphi\right)},
\end{aligned}
\end{equation}
where
\begin{equation}
\begin{aligned}
    E_{C\theta} &= \frac{e^2}{2}\left(\frac{1}{C_{\text{sh}}}+\frac{1}{2C_J}\right),\:\:\: E_{C\phi} = \frac{e^2}{C_J}, \:\:\:E_{C\varphi} = \frac{e^2}{4C_J},\:\:\:  J =\frac{2e^2}{C_J},
\end{aligned}
\end{equation}
when $C_{J}^{(R)}\approx C_J^{(L)}=C_J$.
Figure~\ref{full_circuit_spectra} shows the spectra of the full-circuit model (black dashed), which are obtained by directly diagonalizing Eq.~\eqref{full_circuit_Hamiltonian}, and the spectra of the second-harmonic model (red solid) from Eq.~\eqref{simplified_Hamiltonian}.
We observe that, within our circuit parameter regime, the eigenenergies of the approximate Hamiltonian accurately match those of the the full-circuit Hamiltonian. 
We also simulate the transition spectrum of a more general circuit, which includes an additional inductance outside the SQUID loop, connecting the SQUID and the shunt capacitor. From a conservative choice of the additional inductance ($\approx\SI{22}{pH}$), we find the maximum energy difference between this model and Eq.~\ref{full_circuit_Hamiltonian} is less than the linewidth of the measured transition spectrum.
\newline\newline

\begin{figure}[h]
    \centering
    \subfloat{\label{CJ_dependence_a}}
    \subfloat{\label{CJ_dependence_b}}
    \includegraphics[]{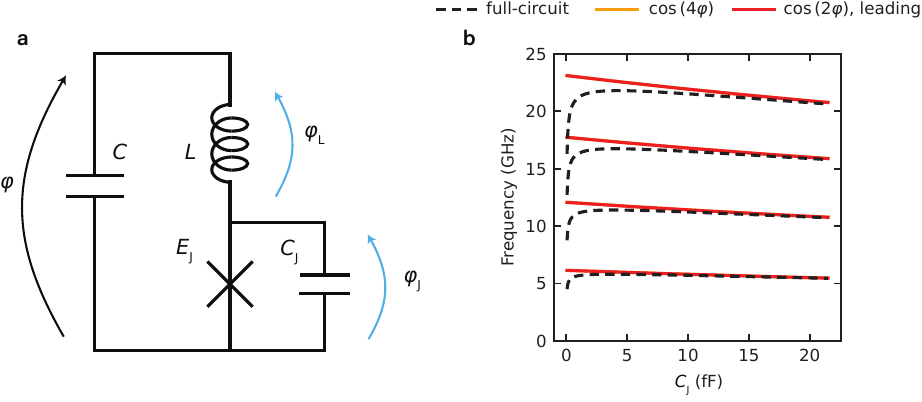}
    \caption{\textbf{(a)} A full-circuit model of a transmon with a series inductance. \textbf{(b)} Differing energies of the second-harmonic approximation in low $C_J$ regime (plot range: $C_J>\SI{0.05}{fF}$). The forth-harmonic approximation result (orange solid) overlaps with the second-harmonic approximation (red solid). Circuit parameters are adopted from Ref.~\cite{Willsch24NatPhy}: $C=\SI{79}{fF}, E_J=\SI{21.8}{GHz}, L=\SI{380}{pH}$.}
    \label{CJ_dependence}
\end{figure}

Moreover, we notice that the current-conservation-based harmonics approximation [Eq.~\eqref{ind_harmonics}] actually requires the junction capacitance to be sufficiently large. 
This becomes clearer when the series inductance is large.
For example, with the transmon parameter regime used in Ref.~\cite{Willsch24NatPhy}, we observe a discrepancy between the second-harmonic approximation and the full-circuit analysis, as shown in Fig.~\ref{CJ_dependence}. In this comparison~[Fig.~\ref{CJ_dependence_b}], we also included the result of the Hamiltonian truncated at the fourth harmonic with higher powers of $E_J/E_L$ in Eq.~\eqref{ind_harmonics} to check possible further corrections.
Here, the Hamiltonian of the full-circuit~[Fig.~\ref{CJ_dependence_a}] is derived from the Lagrangian

\begin{equation}
    L_{\text{full}} = \frac{1}{2}C\dot{\Phi}^2+\frac{1}{2}C_J(\dot{\Phi}-\dot{\Phi}_L)^2+E_J\cos{(\varphi-\varphi_L)}-\frac{1}{2}E_L\varphi_L^2,
\end{equation}
which leads to
\begin{equation}
    H_{\text{full}} = 4(E_C + E_{CJ})\:\hat n_L^2 + \frac{1}{2}E_L\hat\varphi_L^2 + 4E_{C}\:(\hat n-n_g)^2 - E_J\cos{(\hat\varphi-\hat\varphi_L)}+8E_{C}\:(\hat n-n_g)\hat n_L,
\end{equation}
where $E_C=e^2/2C$, $E_{CJ}=e^2/{2C_J}$.
As outlined in the main text, the difference arises as the fluctuation of $\varphi_L$ increases in small $C_J$ regime.

Keeping the discrepancy of the harmonics approximation for the inductance-induced correction in mind, we briefly re-analyze the  fixed-frequency transmon spectra presented in Ref.~\cite{Willsch24NatPhy} (three cooldown data for the KIT sample).
We use two models: 1) the inductance-only model illustrated in Fig.~\ref{CJ_dependence_a} assuming no intrinsic Andreev harmonics, and 2) the same model as 1) but including the second Andreev harmonic $E_{J2, A}$.
For the first model, we optimize $L$, $C_J$, and the resonator coupling strength $J$ as global fit parameters across the three cooldown datasets.
For the second model, we fix $J=85$ MHz and optimize the shunt capacitance $C$ as a global fit parameter to reduce the number of fit variables.
Table.~\ref{KIT_result_1}-~\ref{KIT_result_2} show possible sets of circuit parameters for the two models that agree with the data within 2 MHz.
As such, it is possible that intrinsic harmonics were at play in the measured devices.
However, the large uncertainty (in some cases by more than an order of magnitude) indicates that more data points are needed to obtain statistically reliable parameters. 
For example, this could be done using the SQUID structures as proposed in this work.

\begin{table*}[!htb]
\caption{\label{KIT_result_1} Possible circuit parameter set for transmon spectra presented in Ref.~\cite{Willsch24NatPhy} (KIT) using inductance-only model. Three cooldown data is combined.}
\begin{ruledtabular}
\begin{tabular}{c|ccccccccc}
parameter&
\makecell{$E_{J1}/h$ (cd1)\\ (GHz)}&
\makecell{$E_{J1}/h$ (cd2)\\ (GHz)}&
\makecell{$E_{J1}/h$ (cd3)\\ (GHz)}&
\makecell{$C$ (cd1) \\(fF)}& 
\makecell{$C$ (cd2) \\(fF)}& 
\makecell{$C$ (cd3) \\(fF)}&
\makecell{$C_J$ (fF)}&
$L$ (pH)&
$J$ (MHz)\\
\colrule
value & 25.7 & 24.1 & 15.3 & 79.2 & 78.3 & 77.6 & 1.43 & 644 & 86.8\\
uncertainty & 51.1 & 48.0 & 30.5 & 34.6 & 34.5  & 34.2   & 33.8 & 61.4   & 4.13 \\
\end{tabular}
\end{ruledtabular}
\end{table*}

\begin{table*}[!htb]
\caption{\label{KIT_result_2} Possible circuit parameter set for transmon spectra presented in Ref.~\cite{Willsch24NatPhy} (KIT) using inductance + second Andreev harmonic model. Three cooldown data is combined.}
\begin{ruledtabular}
\begin{tabular}{c|ccccccccc}
parameter&
\makecell{$E_{J1}/h$ (cd1)\\ (GHz)}&
\makecell{$E_{J1}/h$ (cd2)\\ (GHz)}&
\makecell{$E_{J1}/h$ (cd3)\\ (GHz)}&
\makecell{$E_{J2}/h$ (cd1)\\ (GHz)}&
\makecell{$E_{J2}/h$ (cd2)\\ (GHz)}&
\makecell{$E_{J2}/h$ (cd3)\\ (GHz)}&
\makecell{$C$ (fF)}&
\makecell{$L$ (pH)}&
\makecell{$C_J$ (fF)}\\
\colrule
value & 24.8 & 23.5 & 15.0 & 0.277 & 0.193 & 0.0367 & 78.8 & 406 & 1.41\\
uncertainty& 43.9 & 41.4  & 26.5  & 1.75 & 1.22 & 0.251 & 36.9 & 706   & 36.5 \\
\end{tabular}
\end{ruledtabular}
\end{table*}

\newpage
\subsection{Potential in the presence of both sources}
In this section we derive the equation for the total effective second harmonic $E_{J2}\approx E_{J2,A}+E_{J2,L}$ in the presence of both Andreev process and series inductance.
Here, we adopt the method of perturbatively optimizing $\varphi_L$ to minimize the potential discussed in Refs.~\cite{Putterman25Arxiv, Karif17PRA}, which produces the same result as the method used to derive Eq.~\eqref{ind_harmonics_second}.

In the presence of the intrinsic Andreev second harmonic, $E_{J2,A}=\beta E_{J1}$, the potential energy of the series in Fig.~\ref{junction_inductor} is given by
\begin{equation}
\begin{aligned}
    U &= -E_{J1}\Big[\cos(\varphi-\varphi_L)-\beta \cos{(2\varphi - 2\varphi_L)}\Big]+\frac{1}{2}E_L\varphi_L^2\\
    &=-E_{J1}\Big[\cos{(\varphi)}\cos{(\varphi_L)}+\sin{(\varphi)}\sin{(\varphi_L)}\Big]\\
    &\:\:\:\:\:\:\:\:+\beta E_{J1}\Big[\cos{(2\varphi)}\cos{(2\varphi_L)}+\sin{(2\varphi)}\sin{(2\varphi_L)}\Big] + \frac{1}{2}E_L\varphi_L^2,
\end{aligned}
\end{equation}
where $E_{J1}=E_{J1, A}$ is the fundamental harmonic of the Andreev process. 
Expanding the cosines and sines up to the second order of $\varphi_L$, we obtain
\begin{equation}
\begin{aligned}\label{potential_AL}
    U &= -E_{J1}\Big[\cos{(\varphi)}\left(1-\frac{1}{2}\varphi_L^2\right)+\sin{(\varphi)}\:\varphi_L\Big]\\
    &\:\:\:\:\:\:\:\:+\beta E_{J1}\Big[\cos{(2\varphi)}\left(1-2\varphi_L^2\right)+\sin{(2\varphi)}\:(2\varphi_L)\Big] + \frac{1}{2}E_L\varphi_L^2,
\end{aligned}
\end{equation}
which is minimized when
\begin{equation}\label{opt_phiL}
    \varphi_L = \frac{E_{J1}\sin{(\varphi)}-2\beta E_{J1}\sin{(2\varphi)}}{E_L + E_{J1}\cos{\varphi}-4\beta E_{J1}\cos{(2\varphi)}}.
\end{equation}
Again, we expand the right-hand side of the equation using $1/(1-x) \approx 1 + x + x^2 + \cdots$, where the expansion variable $x$ in this case is
\begin{equation}\label{expansion_varialbe}
    x=\frac{E_{J1}}{E_L}\Big(-\cos{(\varphi)}+4\beta\cos{(2\varphi)}\Big).
\end{equation}
Substituting Eq.~\eqref{opt_phiL} and Eq.~\eqref{expansion_varialbe} into Eq.~\eqref{potential_AL}, we obtain the following harmonic expansion.
\begin{equation}
    U = -\sum_{n\geq1} E_{Jn,\text{eff}}\cos{(n\varphi)},
\end{equation}
where
\begin{equation}
\begin{aligned}
    E_{J1\text{,eff}} &= E_{J1} \left[1-\beta\left(\frac{E_{J1}}{E_L}\right)-\frac{1}{8}\left(\frac{E_{J1}}{E_L}\right)^2+\cdots\right]\\
    E_{J2\text{,eff}} &= E_{J1} \left[-\beta-\frac{1}{4} \left(\frac{E_{J1}}{E_L}\right)+\beta\left(\frac{E_{J1}}{E_L}\right)^2+\cdots\right]\\
    E_{J3\text{,eff}} &= E_{J1}\left[\beta\left(\frac{E_{J1}}{E_L}\right)+\frac{1}{8}\left(\frac{E_{J1}}{E_L}\right)^2+\cdots\right].
\end{aligned}
\end{equation}
Similar to the previous sections, truncating the summation at the second harmonic and retaining the leading order with respect to both $\beta$ and ${E_{J1}}/{E_L}$ yields
\begin{equation}
    U \approx -E_{J1}\cos{(\varphi)}+\underbrace{E_{J1}\left(\beta +\frac{E_{J1}}{4E_L}\right)}_{E_{J2\text{, eff}}}\cos{(2\varphi)},
\end{equation}
where the effective second harmonic appears as a sum of the two sources as discussed in the main text.
For notational simplicity, the subscript “eff” will be omitted henceforth.

\newpage
\section{Device transition spectrum measurement} 
\subsection{Fitting algorithms} \label{Section_fitting}
The device Hamiltonian under the second-harmonic approximation is
\begin{equation}
\begin{aligned}
    H &= 4E_C\hat{n}^2 - E_{J1}^{(L)}\cos{\hat\varphi}-E_{J1}^{(R)}\cos{(\hat\varphi-\phi_{\textnormal{ext}})}+E_{J2}^{(L)}\cos{(2\hat\varphi)} +  E_{J2}^{(R)}\cos{\big(2(\hat\varphi-\phi_{\textnormal{ext}})\big)},
\end{aligned}
\end{equation}
where the second harmonic $E_{J2}^{(i)}$ $(i=L,R)$ is given by
\begin{equation}
\begin{aligned}
    E_{J2}^{(i)} &= E_{J2, A}^{(i)} + E_{J2, L}^{(i)} =\beta E_{J1}^{(i)}+\frac{\left(E_{J1}^{(i)}\right)^2}{4E_L}.
\end{aligned}
\end{equation}
Let $\alpha^{(i)}$ denote the ratio between the fundamental and the second harmonics:
\begin{equation}
\begin{aligned}
    \alpha^{(i)} = \frac{E_{J2}^{(i)}}{E_{J1}^{(i)}} = \beta+\frac{E_{J1}^{(i)}}{4E_L}.
\end{aligned}
\end{equation}
Then, we can re-express the Hamiltonian as
\begin{equation}
\begin{aligned}
    \hat{H} &= 4E_C\hat{n}^2 - E_{J1}^{(L)}\Big[\cos{\hat\varphi}-\alpha^{(L)}\cos{(2\hat\varphi)}\Big] - E_{J1}^{(R)}\Big[\cos{(\hat\varphi-\phi_{\textnormal{ext}})}-\alpha^{(R)}\cos{\big(2\hat\varphi-2\phi_{\textnormal{ext}}\big)}\Big].
\end{aligned}
\end{equation}
Additionally, we introduce a new variable, $dE_J$, to represent the asymmetry between the two junctions of the left and right arms:
\begin{equation}
\begin{aligned}
    E_{J1}^{(R)} = E_{J1}^{(L)}\big(1-dE_J\big).
\end{aligned}
\end{equation}
$\alpha^{(L)}$ and $\alpha^{(R)}$ satisfy the following equation
\begin{equation}
\begin{aligned}
    \alpha^{(R)} =\beta+\frac{E_{J1}^{(R)}}{4E_L}=\beta+\frac{E_{J1}^{(L)}-E_{J1}^{(L)}\:dE_J}{4E_L} = \alpha^{(L)}-\frac{E_{J1}^{(L)}}{4E_L}dE_J.
\end{aligned}
\end{equation}
To reduce the number of fit parameters, and given nominally identical junctions ($dE_J\ll1$), we use the same ratio $\alpha$ for both arms:
\begin{equation}
\begin{aligned}
    \alpha^{(R)} \approx \alpha^{(L)} = \alpha,
\end{aligned}
\end{equation}
which yields the fit Hamiltonian
\begin{equation}
\begin{aligned}\label{fit_hamiltonian}
    H_{\textnormal{fit}} &= 4E_C\hat{n}^2 - E_{J1}^{(L)}\Big[\cos{\hat\varphi}-\alpha\cos{(2\hat\varphi)}\Big] - E_{J1}^{(L)}\big(1-dE_J\big)\Big[\cos{(\hat\varphi-\phi_{\textnormal{ext}})}-\alpha\cos{\big(2\hat\varphi-2\phi_{\textnormal{ext}}\big)}\Big].
\end{aligned}
\end{equation}
We optimize the four circuit parameters, $\textbf{x} = \left(E_C, E_{J1}^{(L)}, dE_J, \alpha\right)^T$, by minimizing the following cost function:
\begin{equation}
    \textbf{x}^* = \underset{E_C,\: E_{J1}^{(L)},\:dE_J, \:\alpha  }{\textnormal{argmin}}\:\:\: \sum_{i}\sum_{\phi_{\textnormal{ext}}}\Big| \lambda_i\left(\varphi_\textnormal{ext}; E_C, E_{J1}^{(L)}, dE_J, \alpha\right)-f_i(\varphi_\textnormal{ext})\Big|^2,
\end{equation}
where $\lambda_i$ represents the $i$-th eigenenergy of the Hamiltonian in Eq.~\eqref{fit_hamiltonian}, with the ground state ($i=0$) energy set to zero, and $f_i$ denotes the measured $i$-th transition frequency.

\newpage
\subsection{Details on spectrum measurement}
\begin{figure}[h]
    \centering
    \subfloat{\label{transition_spectra_a}}
    \subfloat{\label{transition_spectra_b}}
    \includegraphics[]{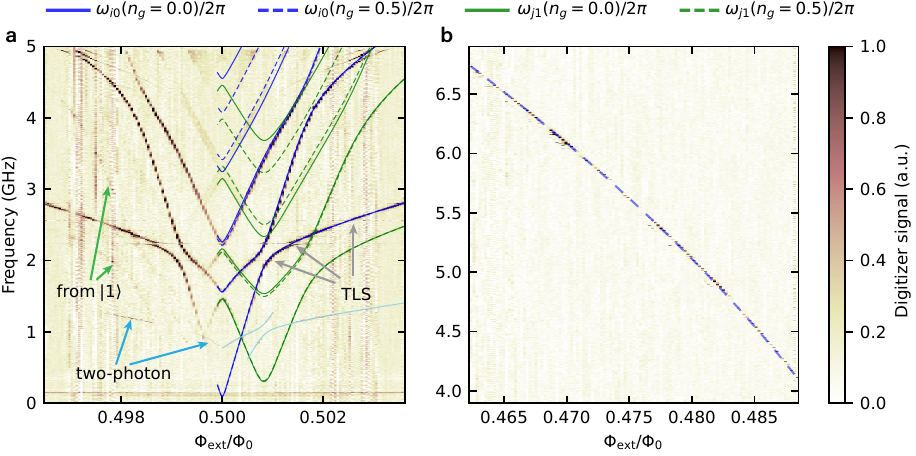}
    \caption{Measured spectra of the $\SI{6}{\micro m}$ device for Chip A \textbf{(a)} near and \textbf{(b)} away from half flux.}
    \label{transition_spectra}
\end{figure}

As shown in Fig.~\ref{transition_spectra} we measure two flux-bias ranges for the fit: one near $\Phi_{\text{ext}}= 0.5\Phi_0$, where the first transition frequency is approximately less than $\SI{3}{GHz}$, and one farther away from $0.5\Phi_0$, where the first transition frequency falls within the $4-7\:\SI{}{GHz}$ range.

\noindent In Fig.~\ref{transition_spectra_a}, additional features apart from the main transitions from the ground state (blue) --- such as the two-photon transition to the first plasmon state (sky blue), transitions from the first excited state (green), and the charge dispersions for both transitions from the ground state and the first excited state (solid and dashed lines)--- are well captured by the second-harmonic model.

\newpage
\section{Effect of qubit-resonator coupling} 
As shown in Section~\ref{Section_fitting}, we fit the measured spectra only using the circuit parameters of the capacitively shunted SQUID.
However, coupling to the readout resonator may induce a frequency shift in the device.
In this section, we discuss the effect of this coupling on the possible frequency shifts.
Given the device Hamiltonian $H_{\textnormal{fit}}$ in Eq.~\eqref{fit_hamiltonian}, the total system Hamiltonian including the resonator is given by
\begin{equation}\label{res_SQUID_hamiltonian}
    H_{\textnormal{sys}} = H_{\textnormal{fit}} + \hbar\omega_r\hat{a}^{\dagger}\hat{a} + \hbar g_c\hat{n}\left(\hat{a}+\hat{a}^{\dagger}\right),
\end{equation}
where $\omega_r$ is the bare resonator frequency, $\hat{a}$ is the annihilation operator of the resonator, and $g_c$ is the coupling strength between the device and the resonator. As discussed in Fig.3 (d)-(f) in the main text, $g_c$ is separately calculated from the resonator single-tone spectroscopy.

\begin{figure}[h]
    \centering
    \subfloat{\label{res_SQUID_spectra_a}}
    \subfloat{\label{res_SQUID_spectra_b}}
    \subfloat{\label{res_SQUID_spectra_c}}
    \subfloat{\label{res_SQUID_spectra_d}}
    \includegraphics[]{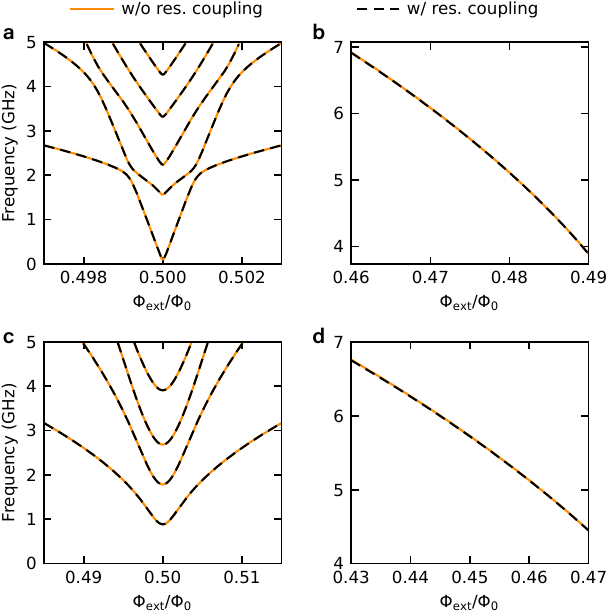}
    \caption{Transition spectrum in the absence (orange solid) and presence (black dashed) of the interaction with the readout resonator. \textbf{(a),(b)} $\SI{6}{\micro m}$ device \textbf{(c),(d)} $\SI{2}{\micro m}$ device for Chip A. We find the frequency shifts of the device due to the interaction are negligible in our experiment.}
    \label{res_SQUID_spectra}
\end{figure}

In Fig.~\ref{res_SQUID_spectra}, we show the transition spectra for the Chip A devices with JJ lengths of $\SI{6}{\micro m}$ (a-b) and $\SI{2}{\micro m}$ (c-d) in the absence (orange solid) and presence (black dashed) of the resonator coupling. 
The former is obtained by diagonalizing Eq.~\eqref{fit_hamiltonian}, and the latter from the system Hamiltonian in Eq.~\eqref{res_SQUID_hamiltonian}.
The energy eigenstates $|\tilde{l}\rangle$ of the system Hamiltonian for Fig.~\ref{res_SQUID_spectra} were chosen based on their maximum overlap with the bare states: $\tilde l = \underset{\tilde k}{\textnormal{argmax}} \big|\langle \tilde k |\left(|0\rangle_{r}\otimes|l\rangle_q\right)\big|$, where $|0\rangle_r$ is the ground state of the bare resonator, and $|l\rangle_q\:\: (l\geq0)$ denotes the $l$th bare eigenstate of the device.
We observe that the coupling strength is weak enough to cause negligible shift in the bare device spectra, indicating that the small coupling to the resonator does not change the fit parameters considerably.

\newpage
\section{Supercurrent diode effect with trace inductance} 
In this section, we discuss supercurrent diode effect estimated in our device. 
In the presence of the second harmonic in the SQUID potential, the time-reversal symmetry can be broken when the device is biased at non-trivial flux $\Phi_\text{ext}\neq n\Phi_0/2$, $n\in\mathbb{Z}$ \cite{Paolucci23APL, Fominov22PRB}.
In this case, the maximum and minimum supercurrents flowing across the SQUID become unbalanced, resulting in the diode effect. 
The effectiveness of the supercurrent diode is often quantified by the rectification efficiency $\eta$ defined as the contrast between the maximum and minimum supercurrents \cite{Paolucci23APL}
\begin{equation}
    \eta = \frac{|I_\text{max}|-|I_\text{min}|}{|I_\text{max}|+|I_\text{min}|}
\end{equation}

\begin{figure}[h]
    \centering
    \subfloat{\label{diode_effect_a}}
    \subfloat{\label{diode_effect_b}}
    \subfloat{\label{diode_effect_c}}
    \includegraphics[]{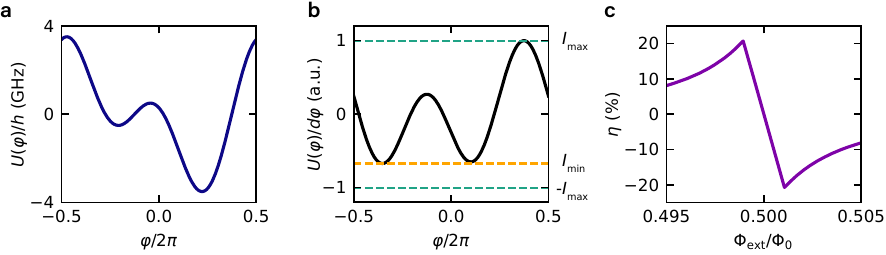}
    \caption{
    \textbf{(a)} Potential energy of the $\SI{6}{\micro m}$ device on Chip A at $\Phi_{\text{ext}}\approx 0.499\:\Phi_0$. 
    \textbf{(b)} Supercurrent across the SQUID estimated from the derivative of the SQUID potential. The difference in $I_\text{max}$ (green dashed) and $I_\text{min}$ (orange dashed) indicate the diode effect.
    \textbf{(c)} Estimated rectification efficiency near half flux. The maximum efficiency is calculated to be over 20\%.
    }
    \label{diode_effect}
\end{figure}

Figure~\ref{diode_effect} shows the diode effect in the $\SI{6}{\micro m}$ device on Chip A. When biased close to half flux, the potential $U(\varphi)$ of the SQUID, calculated from Eq.(3) in the main text, becomes asymmetric due to the second harmonic [Fig.~\ref{diode_effect_a}]. 
The derivative of the potential with respect to the phase $\varphi$ corresponds to the current through the SQUID. 
The difference in magnitudes of the maximum and minimum currents [Fig.~\ref{diode_effect_b}] features the presence of the diode effect in our device.
As shown in Fig.~\ref{diode_effect_c}, our device is estimated to exhibit a rectification efficiency exceeding 10\% near half flux.

Therefore, our results experimentally demonstrate that when engineering the diode effect using series and parallel combinations of JJs, for example, methods proposed in Refs.~\cite{Fominov22PRB, Souto22PRL, Bozkurt23Scipost}, the harmonic corrections arising from the embedding circuitry can be significant and must be carefully considered.
We note that this consideration is not limited to the SIS tunnel junctions used in our work and is generally applicable to different types of JJs, for example, in an all-metallic junction as demonstrated in Ref.~\cite{Paolucci23APL}.

\bibliography{refs}